\documentclass{ws-ijmpe}

\begin{document}

\markboth{JUN XU et al.}{Isospin and momentum dependence of LG
phase transition}

\catchline{}{}{}{}{}
%%%%%%%%%%%%%%%%%%%%%%%%%%%%%%%%%%%%%%%%%%%%%%%%%%%%%%%%%%%%%%%%%%%%

\title{ISOSPIN AND MOMENTUM DEPENDENCE OF LIQUID-GAS PHASE TRANSITION
IN HOT ASYMMETRIC NUCLEAR MATTER}

\author{JUN XU}

\address{Institute of Theoretical Physics, Shanghai Jiao Tong University, Shanghai
200240, China\\
xujuna0307291@sjtu.edu.cn}

\author{LIE-WEN CHEN}

\address{Institute of Theoretical Physics, Shanghai Jiao Tong University, Shanghai
200240, China\\
Center of Theoretical Nuclear Physics, National Laboratory of
Heavy-Ion Accelerator, Lanzhou, 730000, China\\
lwchen@sjtu.edu.cn}

\author{BAO-AN LI}

\address{Department of Physics, Texas A\&M University-Commerce, Commerce, TX
75429-3011, USA\\
Bao-An\_Li@TAMU-Commerce.edu}

\author{HONG-RU MA}

\address{Institute of Theoretical Physics, Shanghai Jiao Tong University, Shanghai
200240, China\\
hrma@sjtu.edu.cn}

\maketitle

\begin{history}
\received{(received date)}
\revised{(revised date)}
%\accepted{(Day Month Year)}
%\comby{(xxxxxxxxxx)}
\end{history}

\begin{abstract}

The liquid-gas phase transition in hot neutron-rich nuclear matter
is investigated within a self-consistent thermal model using
different interactions with or without isospin and/or momentum
dependence. The boundary of the phase-coexistence region is shown
to be sensitive to the density dependence of the nuclear symmetry
energy as well as the isospin and momentum dependence of the
nuclear interaction.

\end{abstract}

\section{Introduction}
\label{introduction}

The liquid-gas (LG) phase transition in nuclear matter remains
illusive and a hot research topic despite of the great efforts
devoted to understanding its nature and experimental
manifestations by the nuclear physics community over many
years\cite{lamb78,fin82,ber83,jaqaman83}. For a recent review,
see, e.g., Refs.\cite{chomaz,das,wci}. Most of the previous
studies have focused on the LG phase transition in symmetric
nuclear matter. While in an asymmetric nuclear matter, the LG
phase transition is expected to display some distinctly new
features because of the isospin degree of freedom and the
associated interactions and additional conservation
laws\cite{muller95}. This expectation together with the need to
understand better properties of asymmetric nuclear matter relevant
for both nuclear physics and
astrophysics have stimulated a lot of new work recently\cite%
{liko97,ma99,wang00,su00,lee01,li01,natowitz02,li02,chomaz03,sil04,lizx04,chomaz06,li07}.
Moreover, the study on the LG phase transition in asymmetric
nuclear matter has received recently a strong boost from the
impressive progress in developing more advanced radioactive beams
that can be used to create transiently in terrestrial laboratories
large volumes of highly asymmetric nuclear matter. Though
significant progress has been made recently in studying properties
of isospin asymmetric nuclear matter and the LG phase transition
in it, there are still many challenging questions to be answered.
Among the main difficulties are our poor understanding about the
isovector nuclear interaction and the density dependence of the
nuclear symmetry energy \cite{wci,ireview98,ibook}. Fortunately,
recent analyses of the isospin diffusion data in heavy-ion
reactions have allowed us to put a stringent constraint on the
symmetry
energy of neutron-rich matter at sub-normal densities \cite%
{betty04,chen05,lichen05}. It is therefore interesting to
investigate how the constrained symmetry energy may allow us to
better understand the LG phase transition in asymmetric nuclear
matter. Moreover, both the isovector (i.e., the nuclear symmetry
potential) and isoscalar parts of the single nucleon potential
should be momentum dependent. However, effects of the
momentum-dependent interactions on the LG phase transition in
asymmetric nuclear matter were not thoroughly investigated
previously.

We report here our recent progress in investigating effects of the
isospin and momentum dependent interactions on the LG phase
transition in hot neutron-rich nuclear matter within a
self-consistent thermal model using three different
interactions\cite{Xu07b}. The first one is the isospin and
momentum dependent MDI interaction constrained by the isospin
diffusion data in heavy-ion collisions. The second one is a
momentum-independent interaction (MID) which leads to a fully
momentum independent single nucleon potential, and the third one
is an isoscalar momentum-dependent interaction (eMDYI) in which
the isoscalar part of the single nucleon potential is momentum
dependent but the isovector part of the single nucleon potential
is momentum independent. We note that the MDI interaction is
realistic, while the other two are only used as references in
order to explore effects of the isospin and momentum dependence of
the nuclear interaction.

\section{Theoretical models}
\label{theory}

\subsection{MDI interaction}
\label{MDI}

In the isospin and momentum-dependent MDI interaction, the potential
energy density $V_{\text{MDI}}(\rho ,T,\delta )$ of a thermally
equilibrated asymmetric nuclear matter at total density $\rho $,
temperature $T$ and isospin asymmetry $\delta $ is expressed as
follows~\cite{das03,chen05},
\begin{eqnarray}
V_{\text{MDI}}(\rho ,T,\delta ) &=&\frac{A_{u}\rho _{n}\rho _{p}}{\rho _{0}}+%
\frac{A_{l}}{2\rho _{0}}(\rho _{n}^{2}+\rho _{p}^{2})+\frac{B}{\sigma +1}%
\frac{\rho ^{\sigma +1}}{\rho _{0}^{\sigma }}(1-x\delta ^{2})  \notag \\
&+&\frac{1}{\rho _{0}}\sum_{\tau ,\tau ^{\prime
}}C_{\tau ,\tau ^{\prime }}\int \int d^{3}pd^{3}p^{\prime }\frac{f_{\tau }(\vec{r},\vec{p}%
)f_{\tau ^{\prime }}(\vec{r},\vec{p}^{\prime
})}{1+(\vec{p}-\vec{p}^{\prime })^{2}/\Lambda ^{2}}. \label{MDIV}
\end{eqnarray}%
In the mean field approximation, Eq. (\ref{MDIV}) leads to the
following
single particle potential for a nucleon with momentum $\vec{p}$ and isospin $%
\tau $ in the thermally equilibrated asymmetric nuclear matter \cite%
{das03,chen05}

\begin{eqnarray}
U_{\text{MDI}}(\rho ,T,\delta ,\vec{p},\tau ) &=&A_{u}(x)\frac{\rho _{-\tau }%
}{\rho _{0}}+A_{l}(x)\frac{\rho _{\tau }}{\rho _{0}}+B(\frac{\rho }{\rho _{0}%
})^{\sigma }  \notag \\
&\times &(1-x\delta ^{2})-8\tau x\frac{B}{\sigma +1}\frac{\rho ^{\sigma -1}}{%
\rho _{0}^{\sigma }}\delta \rho _{-\tau }  \notag \\
&+&\frac{2C_{\tau ,\tau }}{\rho _{0}}\int d^{3}p^{\prime }\frac{f_{\tau }(%
\vec{r},\vec{p}^{\prime })}{1+(\vec{p}-\vec{p}^{\prime
})^{2}/\Lambda ^{2}}
\notag \\
&+&\frac{2C_{\tau ,-\tau }}{\rho _{0}}\int d^{3}p^{\prime }\frac{f_{-\tau }(%
\vec{r},\vec{p}^{\prime })}{1+(\vec{p}-\vec{p}^{\prime
})^{2}/\Lambda ^{2}}. \label{MDIU}
\end{eqnarray}%
In the above the isospin $\tau$ is $1/2$ for neutrons and $-1/2$ for protons, and $f_{\tau }(\vec{%
r},\vec{p})$ is the phase space distribution function at
coordinate $\vec{r}$ and momentum $\vec{p}$. The detailed values
of the parameters $\sigma ,A_{u}(x),A_{l}(x),B,C_{\tau ,\tau
},C_{\tau ,-\tau }$ and $\Lambda $ can be found in Ref.
\cite{das03,chen05} and have been assumed to be temperature
independent here. The isospin and momentum-dependent MDI
interaction gives the binding energy per nucleon of $-16$ MeV,
incompressibility $K_{0}=211$ MeV and the symmetry energy of
$31.6$ MeV for cold symmetric nuclear matter at saturation density
$\rho _{0}=0.16$ fm$^{-3}$ \cite{das03}. The different $x$ values
in the MDI interaction are introduced to vary the density
dependence of the nuclear symmetry energy while keeping other
properties of the nuclear equation of state fixed \cite{chen05}.
We note that the MDI interaction has been extensively used in the
transport model for studying isospin effects in
intermediate-energy heavy-ion collisions induced by
neutron-rich nuclei \cite%
{li04b,chen04,chen05,lichen05,li05pion,li06,yong061,yong062,yong07}.
In particular, the isospin diffusion data from NSCL/MSU have
constrained the value of $x$
to be between $0$ and $-1$ for nuclear matter densities less than about $%
1.2\rho _{0}$ \cite{chen05,lichen05}. We will thus in the present
work consider the two values of $x=0$ and $x=-1$ with $x=0$ giving a
softer symmetry energy while $x=-1$ giving a stiffer symmetry
energy. The potential part of the symmetry energy
$E_{sym}^{pot}(\rho ,x)$ at zero temperature can be parameterized by
\cite{chen05}
\begin{equation}
E_{sym}^{pot}(\rho ,x)=F(x)\frac{\rho }{\rho _{0}}+\left[ 18.6-F(x)\right] (%
\frac{\rho }{\rho _{0}})^{G(x)}, \label{epotsym}
\end{equation}%
where the values of $F(x)$ and $G(x)$ for different $x$ can be found
in Ref.~\cite{chen05}.

\subsection{MID interaction} \label{MID}

In the momentum-independent MID interaction, the potential energy density $%
V_{\text{MID}}(\rho ,\delta )$ of a thermally equilibrated
asymmetric nuclear matter at total density $\rho $ and isospin
asymmetry $\delta $ can be written as
\begin{equation}
V_{\text{MID}}(\rho ,\delta )=\frac{\alpha }{2}\frac{\rho ^{2}}{\rho _{0}}+%
\frac{\beta }{1+\gamma }\frac{\rho ^{1+\gamma }}{{\rho _{0}}^{\gamma }}+{%
\rho }E_{sym}^{pot}(\rho ,x){\delta }^{2}.
\end{equation}%
The parameters $\alpha $, $\beta $ and $\gamma $ are determined by
the incompressibility $K_{0}$ of cold symmetric nuclear matter at
saturation density $\rho _{0}$ as in Ref.~\cite{liko97} and
$K_{0}$ is again set to be $211$ MeV as in the MDI interaction.
And $E_{sym}^{pot}(\rho ,x)$ is just same as Eq.~(\ref{epotsym}).
So the MID interaction reproduces very well the EOS of
isospin-asymmetric nuclear matter with the MDI interaction at zero
temperature for both $x=0$ and $x=-1$. The single nucleon
potential in the MID interaction can be directly obtained as
\begin{equation}
U_{\text{MID}}(\rho ,\delta ,\tau )=\alpha \frac{\rho }{\rho _{0}}+\beta (%
\frac{\rho }{\rho _{0}})^{\gamma }+U^{\text{asy}}(\rho ,\delta ,\tau
),
\end{equation}%
with
\begin{eqnarray}
U^{\text{asy}}(\rho ,\delta ,\tau ) &=&\left[ 4F(x)\frac{\rho }{\rho _{0}}%
+4(18.6-F(x))(\frac{\rho }{\rho _{0}})^{G(x)}\right] {\tau }{\delta
}  \notag
\\
&+&(18.6-F(x))(G(x)-1)(\frac{\rho }{\rho _{0}})^{G(x)}{\delta }^{2}.
\label{Uasy}
\end{eqnarray}%

\subsection{eMDYI interaction}
\label{eMDYI}

The momentum-dependent part in the MDI interaction is also isospin
dependent while the MID interaction is fully momentum independent.
In order to see the effect of the momentum dependence of the
isovector part of the single nucleon potential (nuclear symmetry
potential), we can construct an isoscalar momentum-dependent
interaction, called extended MDYI (eMDYI) interaction since it has
the same functional form as the well-known MDYI interaction for
symmetric nuclear matter \cite{gale90}. In the eMDYI interaction,
the potential energy density $V_{\text{eMDYI}}(\rho ,T,\delta )$
of a thermally equilibrated asymmetric nuclear matter at total
density $\rho $, temperature $T$ and isospin asymmetry $\delta $
is written as
\begin{eqnarray}
V_{\text{eMDYI}}(\rho ,T,\delta ) &=&\frac{A}{2}\frac{\rho ^{2}}{\rho _{0}}+%
\frac{B}{1+\sigma }\frac{\rho ^{1+\sigma }}{{\rho _{0}}^{\sigma }}  \notag \\
&+&\frac{C}{\rho _{0}}\int \int d^{3}pd^{3}p^{\prime }\frac{f_{0}(\vec{r},%
\vec{p})f_{0}(\vec{r},\vec{p}^{\prime })}{1+(\vec{p}-\vec{p}^{\prime
})^{2}/\Lambda ^{2}}+{\rho }E_{sym}^{pot}(\rho ,x){\delta }^{2}.
\label{MDYIV}
\end{eqnarray}%
Here $f_{0}(\vec{r},\vec{p})$ is the phase space distribution
function of
\emph{symmetric nuclear matter} at total density $\rho $ and temperature $T$%
. Again $E_{sym}^{pot}(\rho ,x)$ has the same expression as
Eq.~(\ref{epotsym}). We set $A=\frac{A_{u}+A_{l}}{2}$ and
$C=\frac{C_{\tau ,-\tau }+C_{\tau ,\tau }}{2}$, and $B$, $\sigma $
and $\Lambda $ have the same values as in the MDI interaction, so
that the eMDYI interaction also gives the same EOS of asymmetric
nuclear matter as the MDI interaction at zero temperature for both
$x=0$ and $x=-1$. The single nucleon potential in the eMDYI
interaction can be
obtained as%
\begin{equation}
U_{\text{eMDYI}}(\rho ,T,\delta ,\vec{p},\tau )=U^{0}(\rho ,T,\vec{p}%
)+U^{asy}(\rho ,\delta ,\tau ),
\end{equation}%
where
\begin{eqnarray}
U^{0}(\rho ,T,\vec{p}) &=&A\frac{\rho }{\rho _{0}}+B(\frac{\rho }{\rho _{0}}%
)^{\sigma }+\frac{2C}{\rho _{0}}\int d^{3}p^{\prime }\frac{f_{0}(\vec{r},\vec{p})}{1+(%
\vec{p}-\vec{p}^{\prime })^{2}/\Lambda ^{2}}  \label{Usym}
\end{eqnarray}%
and $U^{\text{asy}}(\rho ,\delta ,\tau )$ is the same as
Eq.~(\ref{Uasy}) which implies that the symmetry potential is
identical for the eMDYI and MID interactions. Therefore, in the
eMDYI interaction, the isoscalar part of the single nucleon
potential is momentum dependent but the nuclear symmetry potential
is not.

\subsection{Thermodynamic Quantities}
\label{thermodynamics}

At zero temperature, $f_{\tau }(\vec{r},\vec{p})$
$=\frac{2}{h^{3}}\Theta (p_{f}(\tau )-p)$ and all the integrals in
above expressions can be calculated analytically \cite{Che07},
while at a finite temperature $T$, the phase space distribution
function becomes the Fermi distribution
\begin{equation}
f_{\tau }(\vec{r},\vec{p})=\frac{2}{h^{3}}\frac{1}{\exp (\frac{\frac{p^{2}}{%
2m_{_{\tau }}}+U_{\tau }-\mu _{\tau }}{T})+1},  \label{f}
\end{equation}%
where $\mu _{\tau }$ is the chemical potential of proton or neutron
and can be determined from%
\begin{equation}
\rho _{\tau }=\int f_{\tau }(\vec{r},\vec{p})d^{3}p.
\end{equation}%
In the above, $m_{_{\tau }}$ is the proton or neutron mass and
$U_{\tau }$ is the proton or neutron single nucleon potential in
different interactions. From a self-consistency iteration scheme
\cite{gale90,xu07}, the chemical
potential $\mu _{\tau }$ and the distribution function $f_{\tau }(\vec{r},%
\vec{p})$ can be determined numerically.

From the chemical potential $\mu _{\tau }$ and the distribution function $%
f_{\tau }(\vec{r},\vec{p})$, the energy per nucleon $E(\rho
,T,\delta )$ can be obtained as
\begin{equation}
E(\rho ,T,\delta )=\frac{1}{\rho }\left[ V(\rho ,T,\delta )+{\sum_{\tau }}%
\int d^{3}p\frac{p^{2}}{2m_{\tau }}f_{\tau }(\vec{r},\vec{p})\right]
. \label{E}
\end{equation}%
Furthermore, we can obtain the entropy per nucleon $S_{\tau }(\rho
,T,\delta )$ as
\begin{equation}
S_{\tau }(\rho ,T,\delta )=-\frac{8\pi }{{\rho
}h^{3}}\int_{0}^{\infty }p^{2}[n_{\tau }\ln n_{\tau }+(1-n_{\tau
})\ln (1-n_{\tau })]dp  \label{S}
\end{equation}%
with the occupation probability%
\begin{equation}
n_{\tau }=\frac{1}{\exp (\frac{\frac{p^{2}}{2m_{_{\tau }}}+U_{\tau
}-\mu _{\tau }}{T})+1}.
\end{equation}%
Finally, the pressure $P(\rho ,T,\delta )$ can be calculated from
the thermodynamic relation
\begin{eqnarray}
P(\rho ,T,\delta ) &=&\left[ T{\sum_{\tau }}S_{\tau }(\rho ,T,\delta
)-E(\rho ,T,\delta )\right] \rho+\sum_{\tau }\mu _{\tau }\rho _{\tau
}.  \label{P}
\end{eqnarray}

\section{LG Phase Transition}
\label{LGphase}

\subsection{Chemical Potential Isobar}
\label{mudelta}

With the above theoretical models, we can now study the LG phase
transition in hot asymmetric nuclear matter. The phase coexistence
is governed by the Gibbs conditions and for the asymmetric nuclear
matter two-phase coexistence equations are
\begin{eqnarray}
\mu _{i}^{L}(T,\rho^{L},\delta^{L}) &=&\mu _{i}^{G}(T,\rho
^{G},\delta^{G}),~(i=n\text{ and }p)  \label{coexistencemu} \\
P^{L}(T,\rho ^{L},\delta^{L}) &=&P^{G}(T,\rho^{G},\delta^{G}),
\label{coexistenceP}
\end{eqnarray}%
where $L$ and $G$ stand for liquid phase and gas phase,
respectively. The chemical stability condition is given by
\begin{equation}
\left( \frac{\partial {\mu }_{n}}{\partial {\delta }}\right)
_{P,T}>0\text{ and }\left( \frac{\partial {\mu }_{p}}{\partial
{\delta }}\right) _{P,T}<0. \label{Cstability}
\end{equation}%
The Gibbs conditions (\ref{coexistencemu}) and (\ref{coexistenceP})
for phase equilibrium require equal pressures and chemical
potentials for two phases with different concentrations and
asymmetries. For a fixed pressure, the two solutions thus form the
edges of a rectangle in the proton and neutron chemical potential
isobars as a function of isospin asymmetry $\delta $ and can be
found by means of the geometrical construction method
\cite{muller95,su00}.

\begin{figure}[htb]
\begin{minipage}{13.5pc}
\includegraphics[scale=0.68]{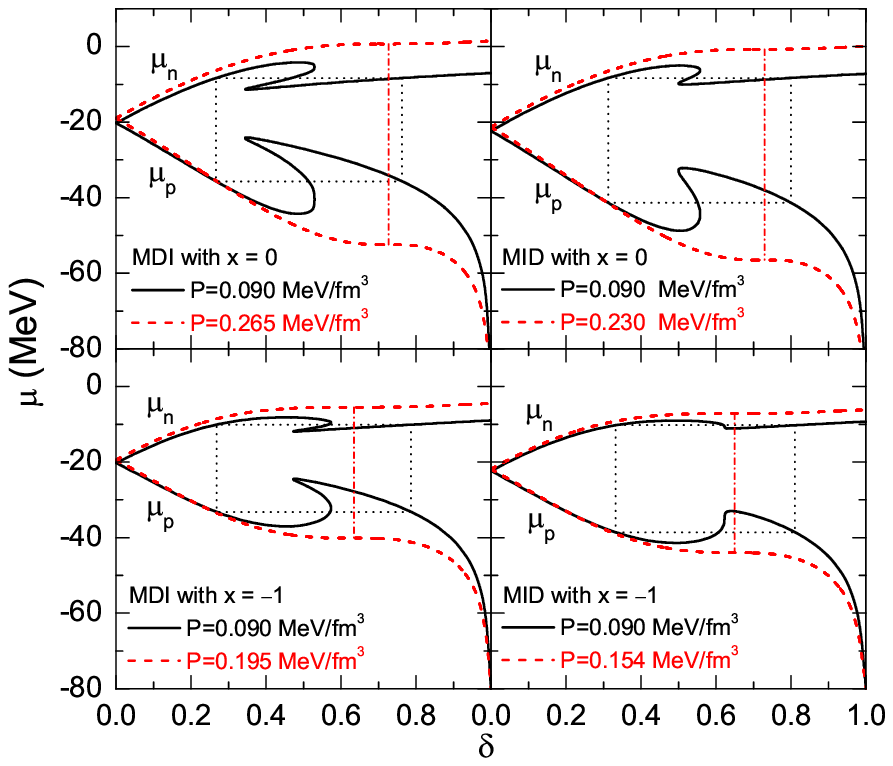}
\end{minipage}
\begin{minipage}{13.5pc}
\includegraphics[scale=0.68]{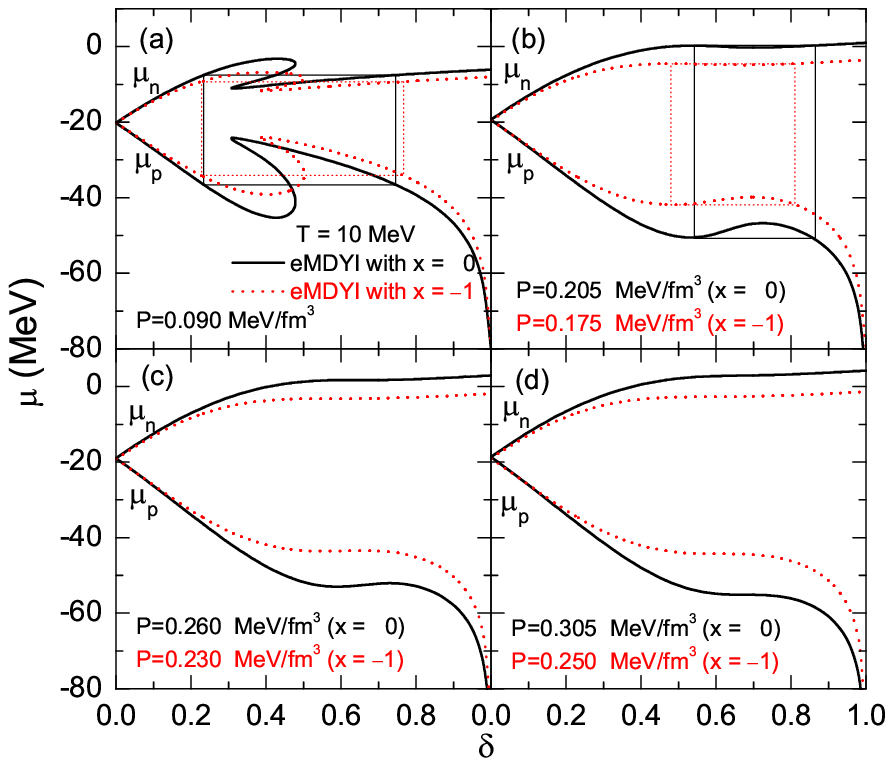}
\end{minipage}
\caption{\protect\small (Color online) The chemical potential isobar
as a function of the isospin asymmetry $\protect\delta $ at $T=10$
MeV for the MDI and MID interactions(left) and the eMDYI
interaction(right) with $x=0$ and $x=-1$. The geometrical
construction used to obtain the isospin asymmetries and chemical
potentials in the two coexisting phases is also shown. Taken from
Ref.~\protect\cite{Xu07b}. } \label{mudelta}
\end{figure}

We calculate the chemical potential isobars at $T=10$ MeV, which
is a typical temperature of LG phase transition. The solid curves
shown in the left panel of Fig.~\ref{mudelta} are the proton and
neutron chemical potential isobars as a function of the isospin
asymmetry $ \delta $ at a fixed temperature $T=10$ MeV and
pressure $P=0.090$ MeV$/$fm$ ^{3}$ by using the MDI and MID
interactions with $x=0$ and $x=-1$. The resulting rectangles from
the geometrical construction are also shown by dotted lines in the
left panel of Fig.~\ref{mudelta}. When the pressure increases and
approaches the critical pressure $P_{\text{C}}$, an inflection
point will appear for proton and neutron chemical potential
isobars. Above the critical pressure, the chemical potential of
neutrons (protons) increases (decreases) monotonically with
$\delta $ and the chemical instability disappears. In the left
panel of Fig.~\ref{mudelta}, we also show the chemical potential
isobar at the critical pressure by the dashed curves. At the
critical pressure, the rectangle is degenerated to a line vertical
to the $\delta $ axis as shown by dash-dotted lines. The values of
the critical pressure are $0.265$, $0.230$, $0.195$ and $ 0.154$
MeV$/$fm$^{3}$ for the MDI interaction with $x=0$, MID interaction
with $x=0$, MDI interaction with $x=-1$ and MID interaction with
$x=-1$, respectively.

Shown in the right panel of Fig.~\ref{mudelta} is the chemical
potential isobar as a function of the isospin asymmetry $\delta $ at
$T=10$ MeV by using the eMDYI interaction with $x=0$ and $x=-1$.
Compared with the results from the MDI and MID interactions, the
main difference is that the left (and right) extrema of $\mu _{n}$
and $\mu _{p}$ do not correspond to the same $\delta $ but they do
for the MDI and MID interactions as shown in the left panel. The
chemical potential of neutrons increases more rapidly with pressure
than that of protons in this temperature. At lower pressures, for
example, $P=0.090$ MeV/fm$^{3}$ as shown in Panel (a), the rectangle
can be accurately constructed and thus the Gibbs conditions
(\ref{coexistencemu}) and (\ref{coexistenceP}) have two solutions.
Due to the asynchronous variation of $\mu _{n}$ and $\mu _{p}$ with
pressure, we will get a limiting pressure $P_{\lim }$ above which no
rectangle can be constructed and the coexistence equations (\ref
{coexistencemu}) and (\ref{coexistenceP}) have no solution. Panel
(b) shows the case at the limiting pressure with $P_{\lim }=0.205$
and $0.175$ MeV/fm$^{3}$ for $x=0$ and $x=-1$, respectively. With
increasing pressure, in Panel (c) $\mu _{n}$ passes through an
inflection point while $\mu _{p}$ still has a chemically unstable
region, and in Panel (d) $\mu _{p}$ passes through an inflection
point while $\mu _{n}$ increases monotonically with $\delta $.

\subsection{Binodal surface}
\label{Pdelta}

For each interaction, the two different values of $\delta $
correspond to two different phases with different densities and the
lower density phase (with larger $\delta $ value) defines a gas
phase while the higher density phase (with smaller $ \delta $ value)
defines a liquid phase. Collecting all such pairs of $\delta (T,P)$
and $\delta ^{\prime }(T,P)$ thus forms the binodal surface.

\begin{figure}[tbh]
\includegraphics[scale=0.7]{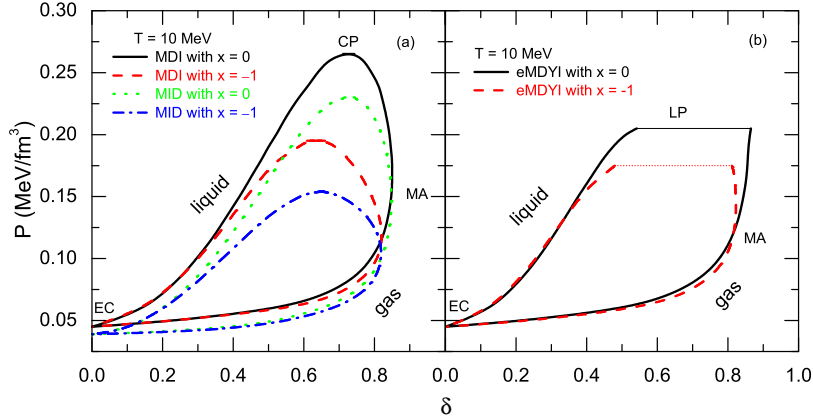}
\caption{{\protect\small (Color online) (a) The section of binodal surface at $%
T=10$ MeV in the MDI and MID interactions with $x=0$ and $x=-1$. The
critical point (CP), the points of equal concentration (EC) and
maximal asymmetry (MA) are also indicated. (b) The section of binodal surface at $%
T=10$ MeV in the eMDYI interaction with $x=0$ and $x=-1$. LP
represents the limiting pressure. Taken from
Ref.~\protect\cite{Xu07b}.}} \label{Pdelta}
\end{figure}

In Fig.~\ref{Pdelta} (a), we show the section of the binodal surface at $%
T=10$ MeV for the MDI and MID interactions with $x=0$ and $x=-1$. On
the left side of the binodal surface there only exists a liquid
phase and on the right side only a gas phase exists. In the region
of \textquotedblleft filet mignon\textquotedblright\ is the
coexistence phase of liquid phase and gas phase. Interestingly, we
can see from Fig.~\ref{Pdelta} (a) that the stiffer symmetry energy
($x=-1$) significantly lowers the critical point (CP) and makes the
maximal asymmetry (MA) a little smaller. Meanwhile, the momentum
dependence in the interaction (MDI) lifts the CP in a larger amount,
while it seems to have no effects on the MA point. In addition, just
as expected, the value of $x$ does not affect the equal
concentration (EC) point while the momentum dependence lifts it
slightly (by about $0.005$ MeV/fm$^{3}$). These features clearly
indicate that the critical pressure and the area of
phase-coexistence region in hot asymmetric nuclear matter is very
sensitive to the stiffness of the symmetry energy with a softer
symmetry energy giving a higher critical pressure and a larger area
of phase-coexistence region. Meanwhile, the critical pressure and
the area of phase-coexistence region are also sensitive to the
momentum dependence. The MDI interaction has a larger area of phase
coexistence region and a larger value of the critical pressure,
compared to the result of MID interaction in the temperature of
$T=10$ MeV.

Fig.~\ref{Pdelta} (b) displays the section of the binodal surface at
$T=10$ MeV by using the eMDYI interaction with $x=0$ and $x=-1$. We
can see that
the curve is cut off at the limiting pressure with $P_{\lim }=0.205$ and $%
0.175$ MeV/fm$^{3}$ for $x=0$ and $x=-1$, respectively. We can also
see that the limiting pressure and the area of phase-coexistence
region are still sensitive to the stiffness of the symmetry energy
with a softer symmetry energy ($x=0$) giving a higher limit pressure
and a larger area of phase-coexistence region in this temperature.

Comparing the results of the MDI and MID interactions shown in
Fig.~\ref{Pdelta} (a), we can see that for pressures lower than the
limiting pressure, the binodal surface from the eMDYI interaction is
similar to that from the MDI interaction. This feature implies that
the momentum dependence of the symmetry potential has little
influence on the LG phase transition in hot asymmetric nuclear
matter while the momentum dependence of the isoscalar single nucleon
potential significantly changes the area of phase-coexistence region
for pressures lower than the limiting pressure. For pressures above
the limiting pressure, the momentum dependence of both the isoscalar
and isovector single nucleon potentials becomes important.

\section{Summary}
\label{summary}

In summary, we have studied the liquid-gas phase transition in hot
neutron-rich nuclear matter within a self-consistent thermal model
using three different nuclear effective interactions, namely, the
isospin and momentum dependent MDI interaction constrained by the
isospin diffusion data in heavy-ion collisions, the
momentum-independent MID interaction, and the isoscalar
momentum-dependent eMDYI interaction. At zero temperature, the
above three interactions give the same EOS for asymmetric nuclear
matter. By analyzing liquid-gas phase transition in hot
neutron-rich nuclear matter with the above three interactions, we
find that the boundary of the phase-coexistence region is very
sensitive to the density dependence of the nuclear symmetry
energy. A softer symmetry energy leads to a higher critical
pressure and a larger area of the phase-coexistence region. In
addition, the area of phase-coexistence region are also seen to be
sensitive to the isospin and momentum dependence of the nuclear
interaction. For the isoscalar momentum-dependent eMDYI
interaction, a limiting pressure above which the liquid-gas phase
transition cannot take place has been found.

\section{Acknowledgements}
\label{acknowledgements}

This work was supported in part by the National Natural Science
Foundation of China under Grant Nos. 10334020, 10575071, and
10675082, MOE of China under project NCET-05-0392, Shanghai
Rising-Star Program under Grant No. 06QA14024, the SRF for ROCS,
SEM of China, the China Major State Basic Research Development
Program under Contract No. 2007CB815004, the US National Science
Foundation under Grant No. PHY-0652548 and the Research
Corporation under Award No. 7123.


\begin{thebibliography}{99}
\bibitem{lamb78} D. Q. Lamb, J. M. Lattimer, C. J. Pethick, and D.
G. Ravenhall, Phys. Rev. Lett. \textbf{41} (1978) 1623.

\bibitem{fin82} J.E. Finn \textit{et al.}, Phys. Rev. Lett. \textbf{49}
(1982) 1321.

\bibitem{ber83} G.F. Bertsch and P.J. Siemens, Phys. Lett. \textbf{%
B126} (1983) 9.

\bibitem{jaqaman83} H. Jaqaman, A. Z. Mekjian, and L. Zamick, Phys.
Rev. C \textbf{27} (1983) 2782; \textit{ibid}. C \textbf{29}
(1984) 2067.

\bibitem{chomaz} Ph. Chomaz, M. Colonna, and J. Randrup, Phys. Rep.
\textbf{389} (2004) 263.

\bibitem{das} C.B. Das, S. Das Gupta, W.G. Lynch, A.Z. Mekjian, and
M.B. Tsang, Phys. Rep. \textbf{406} (2005) 1.

\bibitem{wci} \textit{Dynamics and Thermodynamics with Nucleonic
Degrees of Freedom}, Eds. Ph. Chomaz, F. Gulminelli, W. Trautmann
and S.J. Yennello, Springer, (2006).

\bibitem{muller95} H. M\"{u}ller and B.D. Serot, Phys. Rev. C
\textbf{52} (1995) 2072.

\bibitem{liko97} B.A. Li and C.M. Ko, Nucl. Phys. \textbf{A618} (1997) 498.

\bibitem{ma99} Y.G. Ma \textit{et al.}, Phys. Rev. C \textbf{60} (1999) 024607.

\bibitem{wang00} P. Wang, Phys. Rev. C \textbf{61} (2000) 054904.

\bibitem{su00} W.L. Qian, R.K. Su, and P. Wang, Phys. Lett. \textbf{%
B491} (2000) 90.

\bibitem{lee01} S.J. Lee and A. Z. Mekjian, Phys. Rev. C \textbf{63}%
 (2001) 044605.

\bibitem{li01} B.A. Li, A.T. Sustich, M. Tilley, and B. Zhang, Phys.
Rev. C \textbf{64} (2001) 051303(R)

\bibitem{natowitz02} J.B. Natowitz et al., Phys. Rev. Lett. \textbf{%
89} (2002) 212701.

\bibitem{li02} B.A. Li, A.T. Sustich, M. Tilley, and B. Zhang, Nucl.
Phys. \textbf{A699} (2002) 493.

\bibitem{chomaz03} P. Chomaz and J. Margueron, Nucl. Phys. \textbf{%
A722} (2003) 315c; J. Margueron and P. Chomaz, Phys. Rev. C
\textbf{67} (2003) 041602(R).

\bibitem{sil04} T. Sil, S.K. Samaddar, J.N. De, and S. Shlomo, Phys.
Rev. C \textbf{69} (2004) 014602.

\bibitem{lizx04} Z. Li and M. Liu, Phys. Rev. C \textbf{69} (2004) 034615.

\bibitem{chomaz06} C. Ducoin, P. Chomaz, and F. Gulminelli, Nucl.
Phys. \textbf{A771} (2006) 68; \textit{ibid}. \textbf{A781} (2007)
407.

\bibitem{li07} B.A. Li, L.W. Chen, H.R. Ma, J. Xu, and G.C. Yong,
arXiv:0710.2877 [nucl-th], PRC, in press.

\bibitem{ireview98} B.A. Li, C.M. Ko, and W. Bauer, topical review,
Int. Jour. Mod. Phys. E \textbf{7} (1998) 147.

\bibitem{ibook} \textit{Isospin Physics in Heavy-Ion Collisions at
Intermediate Energies}, Eds.. Bao-An Li and W. Udo Schr\"{o}der
(Nova Science Publishers, Inc, New York, 2001).

\bibitem{betty04} M.B. Tsang \textit{et al.}, Phys. Rev. Lett.
\textbf{92} (2004) 062701.

\bibitem{chen05} L.W. Chen, C.M. Ko, and B.A. Li, Phys. Rev. Lett.
\textbf{94} (2005) 032701 [arXiv:nucl-th/0407032].

\bibitem{lichen05} B.A. Li and L.W. Chen, Phys. Rev. C \textbf{72} (2005)
064611.

\bibitem{Xu07b} J. Xu, L.W. Chen, B.A. Li and H.R. Ma, Phys. Lett. \textbf{%
B650} (2007) 348; arXiv:0710.5409 [nucl-th].

\bibitem{das03} C. B. Das, S. Das Gupta, C. Gale and B.A. Li, Phys.
Rev. C \textbf{67} (2003) 034611.

\bibitem{li04b} B.A. Li, C. B. Das, S. Das Gupta, and C. Gale, Phys.
Rev. C \textbf{69} (2004) 011603(R); Nucl. Phys. \textbf{A735}
(2004) 563.

\bibitem{chen04} L.W. Chen, C.M. Ko, and B.A. Li, Phys. Rev. C
\textbf{69} (2004) 054606.

\bibitem{li05pion} B.A. Li, G.C. Yong, and W. Zuo, Phys. Rev. C
\textbf{71} (2005) 014608; \textit{ibid}. C \textbf{71} (2005)
044604.

\bibitem{li06} B.A. Li, L.W. Chen, G.C. Yong, and W. Zuo, Phys.
Lett. \textbf{B634} (2006) 378.

\bibitem{yong061} G.C. Yong, B.A. Li, L.W. Chen, and W. Zuo, Phys.
Rev. C \textbf{73} (2006) 034603.

\bibitem{yong062} G.C. Yong, B.A. Li, and L.W. Chen, Phys. Rev. C
\textbf{74} (2006) 064617 [arXiv:nucl-th/0606003].

\bibitem{yong07} G.C. Yong, B.A. Li, and L.W. Chen, Phys. Lett.
\textbf{B650} (2007) 344.

\bibitem{gale90} C. Gale, G.M. Welke, M. Prakash, S.J. Lee, and S.
Das Gupta, Phys. Rev. C \textbf{41} (1990) 1545.

\bibitem{Che07} L.W. Chen, C.M. Ko and  B.A. Li, arXiv:0709.0900 [nucl-th],
PRC, (2007) in press.

\bibitem{xu07} J. Xu, L.W. Chen, B.A. Li and H.R. Ma, Phys. Rev. C
\textbf{75} (2007) 014607.
\end{thebibliography}
\end{document}